\RequirePackage{lineno}
\documentclass[aps,prl,twocolumn,showpacs,preprintnumbers,amsmath,amssymb,floatfix,superscriptaddress,lashbox,nofootinbib,groupedaddress]{revtex4}  % for review and submission
\usepackage{pict2e}
\usepackage{graphicx}  % needed for figures
\usepackage{dcolumn}   % needed for some tables
\usepackage{bm}        % for math
\usepackage{amssymb}   % for math
\usepackage{color}

\begin{document}
\widetext
%\leftline{Version 15  as of \today}
%\leftline{Primary authors:WF,PLG,AM and GCT}
%leftline{To be submitted to PRL}
%%\centerline{LVD INTERNAL DOCUMENT -- NOT FOR PUBLIC DISTRIBUTION}
\setpagewiselinenumbers

\title{Measurement of the velocity of neutrinos from the CNGS beam with the Large Volume Detector}

%\author{The LVD Collaboration}
%\today

%\date{\today}
%\maketitle

%\begin{center}
\author{N.Yu.~Agafonova}
\affiliation{Institute for Nuclear Research, Russian Academy of Sciences, Moscow, 117312 Russia}
\author{M.~Aglietta}
\affiliation {INFN-Torino, OATO-Torino, 10100 Torino, Italy} 
\author{P.~Antonioli}
\affiliation {INFN-Bologna, 40126 Bologna, Italy}
\author{V.V.~Ashikhmin}
\affiliation{Institute for Nuclear Research, Russian Academy of Sciences, Moscow, 117312 Russia}
\author{G.~Bari}
\affiliation {INFN-Bologna, 40126 Bologna, Italy}
\author{R.~Bertoni}
\affiliation {INFN-Torino, OATO-Torino, 10100 Torino, Italy} 
\author{E.~Bressan}
\affiliation {University of Bologna, 40126 Bologna, Italy}
\affiliation{Centro Enrico Fermi, 00184 Roma, Italy}
\author{G.~Bruno}
\affiliation {INFN, Laboratori Nazionali del Gran Sasso, 67100 Assergi L'Aquila, Italy}
\author{V.L.~Dadykin}
\affiliation{Institute for Nuclear Research, Russian Academy of Sciences, Moscow, 117312 Russia}
\author{W.~Fulgione}
\affiliation {INFN-Torino, OATO-Torino, 10100 Torino, Italy} 
\author{P.~Galeotti}
\affiliation{University of Torino, 10125 Torino, Italy}
\affiliation {INFN-Torino, OATO-Torino, 10100 Torino, Italy} 
\author{M.~Garbini}
\affiliation {INFN-Bologna, 40126 Bologna, Italy}
\affiliation {INFN-Bologna, 40126 Bologna, Italy}
\author{P.L.~Ghia}
\affiliation{Laboratoire de Physique Nucl\'{e}aire et de Hautes Energies (LPNHE), Universit\'{e}s 
Paris 6 et Paris 7, CNRS-IN2P3, Paris,  France}
\author{P.~Giusti}
\affiliation {INFN-Bologna, 40126 Bologna, Italy}
\author{E.~Kemp}
\affiliation {University of Campinas, 13083-859 Campinas, SP, Brazil}
\author{A.S.~Mal'gin}
\affiliation{Institute for Nuclear Research, Russian Academy of Sciences, Moscow, 117312 Russia}
\author{B.~Miguez}
\affiliation {University of Campinas, 13083-859 Campinas, SP, Brazil}
\affiliation {INFN, Laboratori Nazionali del Gran Sasso, 67100 Assergi L'Aquila, Italy}
\author{A.~Molinario}
\affiliation {INFN-Torino, OATO-Torino, 10100 Torino, Italy} 
\author{R.~Persiani}
\affiliation {INFN-Bologna, 40126 Bologna, Italy}
\affiliation {University of Bologna, 40126 Bologna, Italy}
\author{I.A.~Pless}
\affiliation{Massachusetts Institute of Technology, Cambridge, MA 02139-4307, USA}
\author{V.G.~Ryasny}
\affiliation{Institute for Nuclear Research, Russian Academy of Sciences, Moscow, 117312 Russia}
\author{O.G.~Ryazhskaya}
\affiliation{Institute for Nuclear Research, Russian Academy of Sciences, Moscow, 117312 Russia}
\author{O.~Saavedra}
\affiliation{University of Torino, 10125 Torino, Italy}
\affiliation {INFN-Torino, OATO-Torino, 10100 Torino, Italy} 
\author{G.~Sartorelli}
\affiliation {INFN-Bologna, 40126 Bologna, Italy}
\affiliation {University of Bologna, 40126 Bologna, Italy}
\author{I.R.~Shakyrianova}
\affiliation{Institute for Nuclear Research, Russian Academy of Sciences, Moscow, 117312 Russia}
\author{M.~Selvi}
\affiliation {INFN-Bologna, 40126 Bologna, Italy}
\author{G.C.~Trinchero}
\affiliation {INFN-Torino, OATO-Torino, 10100 Torino, Italy} 
\author{C.~Vigorito}
\affiliation{University of Torino, 10125 Torino, Italy}
\affiliation {INFN-Torino, OATO-Torino, 10100 Torino, Italy} 
\author{V.F.~Yakushev}
\affiliation{Institute for Nuclear Research, Russian Academy of Sciences, Moscow, 117312 Russia}
\author{A.~Zichichi}
\affiliation {INFN-Bologna, 40126 Bologna, Italy}
\affiliation {University of Bologna, 40126 Bologna, Italy}
\affiliation{Centro Enrico Fermi, 00184 Roma, Italy}
\affiliation {CERN, Geneva, Switzerland}
\author{A.~Razeto}
\affiliation {INFN, Laboratori Nazionali del Gran Sasso, 67100 Assergi L'Aquila, Italy}
\collaboration {The LVD Collaboration}
\noaffiliation
%\maketitle
\begin{abstract}
We report the measurement of the time-of-flight of $\sim$ 17 GeV $\nu_{\mu}$ on the CNGS baseline (732 km) with the Large Volume Detector (LVD) at the Gran Sasso Laboratory.
The CERN-SPS accelerator has been operated from May 10th to May 24th 2012,with a tightly bunched-beam structure to allow the velocity of neutrinos to be accurately measured on an event-by-event basis.
LVD has detected 48 neutrino events, associated to the beam, with a high absolute time accuracy. 
These events allow to establish the following limit on the difference between the neutrino speed and the light velocity:  $-3.8\cdot 10^{-6} < (v_{\nu} - c)/c < 3.1 \cdot 10^{-6}$ (at 99\% C.L.). This value is an order of magnitude lower than previous direct measurements.

\end{abstract}
\date{\today}
\maketitle
%\linenumbers
%\pagebreak

{\it{Introduction.-}}
Cosmological measurements \cite{cosmological} provide stringent limits on the sum of neutrino masses. 
Even assuming the heaviest neutrino eigenstate, the expected relative deviation from speed of light is lower than $10^{-20}$ for $\approx$ 10 GeV neutrinos.
Nevertheless, in the past, theories allowing some or all neutrinos to have apparent velocities different than  the speed of light, $c$, (see e.g. \cite{Pas}) have been proposed.
A stringent limit at $E\approx10$ MeV, $|v_{\nu} - c|/c<2\times 10^{-9}$, has been obtained from the observation of SN1987A electron anti-neutrinos~\cite{SN1987A}.
At higher energies (E $>$ 30 GeV), the deviation has been tested down to $|v_{\nu}-c|/c < 4 \times 10^{-5}$ \cite{PRL}. 
The MINOS collaboration \cite{[minos]} has performed a neutrino time-of-flight ($tof$) measurement on a $\approx$ 735 km baseline and with a beam with average energy $<E>$=3 GeV. The MINOS result is quoting $(v{_\nu} - c)/c = (5.1 \pm 2.9)\times 10^{-5}$ at 68\% C.L.. 
Recently the OPERA collaboration has reported \cite{opera} a $tof$ measurement for muon neutrinos from the CNGS (CERN Neutrinos to Gran Sasso) beam. The reported evidence of a superluminal propagation of $\nu_{\mu}$ was subsequently attributed to a technical problem \cite{Zic}.
%\cite{opera-lvd}\cite{opera-disc}.
This explanation has been also confirmed by the results of another experiment in the Gran Sasso Laboratory \cite{icarus}.\\
The Large Volume Detector (LVD), in the INFN Gran Sasso National Laboratory (LNGS), at the average depth of 3600 m w.~e., is a 1 kt liquid scintillator detector whose major purpose is monitoring the Galaxy to study neutrino bursts from gravitational stellar collapses \cite{LVD}. It has started operation in 1992, and since 2006 it has been acting as a far-monitor of the CNGS beam \cite{LVDmonitor}\cite{cngsRun1}. 
LVD is sensitive to neutrino interactions with protons and carbon nuclei in the liquid scintillator and with the iron of the detector structure. Muons, produced by charged current interactions of muon neutrinos in the rock, can also be detected and are responsible for the bulk of CNGS events in LVD. 
We report a measurement of the neutrino velocity obtained through the detection of $\nu_{\mu}$ from the CNGS beam. We show that the neutrino speed is compatible with $c$, its deviation being  
$< 3.8 \times 10^{-6}$ 
at 99\% C.L.. The sensitivity of this measurement is by one order of magnitude better than that of previous ones in a similar energy range.\\
{\it{The detector. -}} LVD consists of 840 scintillator counters, 1.5  m$^3$ each. 
The array is divided in three identical ``towers''  with independent high voltage power supply, trigger, data acquisition and absolute clock (ESAT Slave RAD100) connected, through a 8 km optical link, to the Master clock (ESAT RAD100) located in the external buildings of the LNGS.
Each tower consists of 35 ``modules'' hosting a cluster of 8 counters.
Each counter is viewed from the top by three 15 cm photomultiplier tubes (PMTs) FEU49b or FEU125.\\
LVD standard electronics is described in detail in \cite{Bigongiari}\cite{fulg}.
The trigger condition  for each tower is the three-fold coincidence of the PMTs of  any of its counters, corresponding, on average, to the energy threshold ${\cal E}_H \simeq 4$ MeV.
The energy released in every counter is measured, through 12 bit ADCs, with a mean resolution $\sigma_{E}/E \sim $ 15\% at 10 MeV.
The arrival time is measured with a 12.5 ns granularity.
A high stability Citrine crystal (40 MHz) oscillator supplies the general clock for the whole experiment. 
%
%
%
%
%{One millisecond after the occurrence of a trigger, the memory buﬀers, containing the energy and time information of all detected signals, are read out together with the associated time stamps from the clock (ESAT Slave) of each tower. Due to the 100 ns granularity of UTC clock, the intrinsic precision is of the order of 100 ns/$\sqrt(12) \approx$ 30 ns. 
%
One millisecond after the occurrence of a trigger, the memory buffers, containing the energy and time information of all detected signals, are read out together with the associated time stamp from every tower's absolute clock (ESAT Slave).
The absolute time accuracy of LVD is of the order of one microsecond.
This is better than what needed to search for coincidences among different neutrino telescopes generated by a gravitational collapse and to guide the search for possible gravitational-wave signals \cite{Pagliaroli}. \\
In spite of its limited absolute time resolution, LVD has detected, at the end of 2011, 32 neutrino events from the CNGS bunched-beam test \cite{CERN1}.
%\footnote{The CNGS bunched-beam test lasted from October 22 to November 6, 2011 for a total integrated intensity of 4$\cdot$10$^{16}$ protons on target.}. 
The difference between the neutrino time-of-flight {\it tof}$_\nu$ and the expected one at speed of light {\it tof}$_c$ has been found to be:
\par
\begin{figure}[htb]
%\centering
\vspace{-.0cm}
\includegraphics[width=90mm] {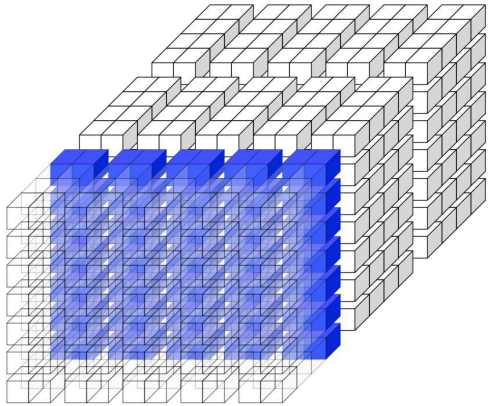}
\vspace{-.0cm}
\caption{Schema of LVD counters. The darkened (blue) ones represent the Super-Set, see text.}
\label{fig:0c}
\end{figure}
\begin{figure}[htb]
%\centering
\vspace{-.0cm}
\includegraphics[width=90mm] {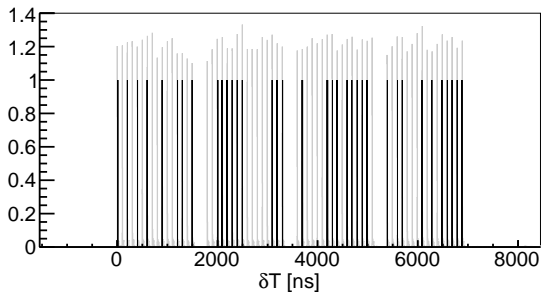}
\vspace{-.0cm}
\caption{Comparison of the $\delta T$ values of the 48 selected events (black lines) with the summed waveforms of proton extraction (grey lines). The origin of time for the waveforms is given by the maximum of the first bunch.}
\label{fig:wfdistr}
\end{figure}
\begin{figure}[htb]
%\centering
\vspace{-.0cm}
\includegraphics[width=90mm] {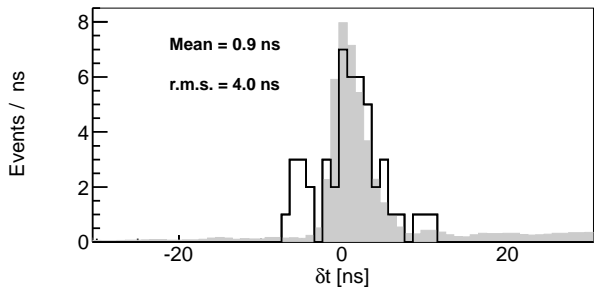}
\vspace{-.0cm}
\caption{Distribution of $\delta t$, the difference between the time-of-flight of neutrino and the time-of-flight at speed of light, for the 48 selected events (black histogram), compared with the superposition of the peaks of the waveforms correlated to detected
events (grey histogram).}
\label{fig:finaldistr}
\end{figure}

\vspace{-.5cm}
\begin{equation}\label{delta_2011}
  \delta t={\it tof}_\nu-{\it tof}_c = 3.1\pm 5.3_{stat} \pm 8_{syst} ~ {\rm ns}
\end{equation}
To perform this measurement, we have carefully re-calibrated all LVD components and time delays in order to improve time accuracy. 
The uncertainties are due to the absolute time resolution; to the fluctuations in the time response of the different detectors (average values of transit times were used when not known); and to the uncertainty in determining the event position. 
This preliminary measurement has shown that, within the uncertainties, the speed of CNGS neutrinos is compatible with that of the light. Moreover, it has helped identifying the detector characteristics to be improved for a more accurate measurement of the neutrino velocity.\\
{\it{Setup upgrades. - }} To allow a very accurate neutrino $tof$ measurement, from May 10$^{th}$ to May 24$^{th}$, 2012, the CERN-SPS accelerator has been operated with a new beam structure.
This structure was made of four batches separated by about 300 ns, with 16 bunches per batch, with a narrow width of $\sim$ 3 ns each, separated by 100 ns.
The batched structure is known through the waveform obtained from the Beam Current Transformer (BCT) (see fig. \ref{fig:wfdistr}, grey lines). \\
For this occasion, a new High Precision Time Facility (HPTF) has been designed by the Borexino collaboration \cite{Alessandro} and installed in the external buildings of the LNGS. 
%
%
%
%A Septentrio PolaRx4 clock provides the GPS  time, PPS (1 Hz) and XPPS (10 Hz) output signals. 
A Septentrio PolaRx4 GNSS receiver, synchronized with the 10 MHz frequency of a GPS disciplinated Rubidium clock, provides GPS time and XPPS (10 Hz) output signal.
The HPTF is equipped with high precision (50 ps) Time Interval Counters (TIC) Pendulum CNT-91, to which the triggers of the different LNGS experiments can be connected.\\
In view of the new  neutrino beam, we have modified a subset of the LVD counters, to improve their timing performances. We have chosen 58 of them (Super-Set, see fig. \ref{fig:0c}) to maximize the acceptance with respect to CNGS neutrinos while minimizing the number of detectors to be modified. 
From data taken since 2006 from the CNGS beam 
%(in excess of $10^{5}$ events) 
(and in agreement with Monte Carlo simulations \cite{LVDmonitor})
we have measured that the Super-Set counters are involved in $\sim\,$40\% of the CNGS events detected by LVD, while representing only less than 7\% of the whole array.
On one hand, to avoid time fluctuations in the trigger formation at the single-counter level, we have modified the cabling of the PMTs, by delaying only one of them. The change guarantees that the 3-fold coincidence among the PMTs in every counter is always formed due to the same tube. On the other hand, to perform a measurement of the transit time in each counter, we have equipped them with a LED system.
The transit time, denominated $\delta_{LVD}$, is the time between the light generation inside a counter and the formation of the trigger.
By means of the LED system, we have measured the behavior of the transit time versus energy for each counter. 
By varying the LED intensity we have simulated different energy releases in each of them, and measured the transit time versus energy. 
For each counter, we have best fitted the results with a power-law of the form:
\begin{equation}
\delta_{LVD}\,(E_{m})\,=\,P_{1}\,\cdot\,E_{m}^{P_{2}}\,+\,k
\end{equation}
where E$_{m}$ is the energy release measured by the ADC, while P1, P2 and k are free parameters of the fit, for each counter.\\
The trigger of the array has been upgraded too. An independent fast trigger logic has been implemented by extracting the coincidence of the triggered counters from discriminators, with the trigger sent to the HPTF. The trigger has been connected to one of the TIC, this being stopped by the XPPS signal. This provides a high precision time difference, $\delta_{TIC}$, between the LVD trigger and the absolute GPS time. Thanks to this improvement, the absolute time accuracy of LVD is of the order of few nanoseconds.\\
Finally a new, indipendent, high precision geodesy measurement has been performed. The CNGS-LVD baseline, namely  the distance between the center of the BFCTI.400344 Beam Current Transformer (BCT) detector at CERN and the LVD Super-Set upstream entry wall (taken as the LVD reference), is found to be 731291.87$\pm$0.04~m\cite{Barzaghi}. 
The corresponding time-of-flight at speed of light, when including the 2.2 ns contribution due to Earth's rotation, is $tof_{c}\,=\,2439329.32\pm0.13\,$ns. \\
{\it{The Measurement.-}} The time-of-flight of neutrinos, $tof_{\nu}$, is the difference between the absolute time at which the event triggers a counter in LVD ($t_{\nu}$) and the absolute time at which the proton bunch crosses the BCT intensity monitor
at CERN ($t_{p}$), both expressed in the LNGS reference time, i.e., $tof_{\nu}=t_{\nu}-t_{p}$.\\
$t_{\nu}$ is given by:
\begin{equation}\label{tnu}
%t_{\nu}\,=\,t^{\rm GPS}_{\rm XPPS}\,-\delta_{\rm TIC}\,-\,\delta_{LVD}\,-\,\delta_{h}\,+\,\Delta_{INRIM}
%t_{\nu}\,=\,t_{gps}\,-\,TIC\,-\,\delta_{LVD}\,-\,\delta_{h}\,+\,4.5\,{\rm ns}
t_{\nu}\,=\,t^{\rm GPS}_{\rm XPPS}\,-\delta_{\rm TIC}\,-\,\delta_{LVD}\,-\,\delta_{h}\,+\,\Delta_{const}^{LVD}
\end{equation}
$t^{\rm GPS}_{\rm XPPS} $ is the absolute GPS time stamp of the first XPPS signal after the trigger. 
$\delta_{h}$ is a correction for the path difference due to neutrino events hitting counters at different heights (the angle of incidence of the CNGS beam with respect to the horizontal plane is $\approx$ 3 deg). 
$\Delta_{const}^{LVD} = 180.8 \pm 1.5$ ns sums up all constant delays in LVD. 
$\delta_{LVD}$ and $\delta_{\rm TIC}$ have been previously introduced. 
They account for the transit time correction of each counter and the delay between the trigger and the absolute GPS time, respectively.\\
$t_{p}$ is given by:
\begin{equation}
t_{p}\,=\,t_{HCA}\,+\,\delta_{wf}\,+\,\Delta_{const}^{CERN}\,+\,\delta_{LNGS}^{CERN}
\end{equation}
where $t_{HCA}$ is the time stamp of the kicker proton extraction signal
(start of the BCT waveform digitalization); $\delta_{wf}$ is the delay of the
first proton bunch with respect to the beginning of the waveform
acquisition ($t_{HCA}$);
$\Delta_{const}^{CERN}\,=\,9521.1\pm 2.0$ ns sums up all the constant delays
at CERN and $\delta_{LNGS}^{CERN}$ is the time difference between the LNGS and CERN reference systems \cite{Alessandro}.\\
LVD has been fully operational during the May 2012 bunched-beam run.
For each detected event in that period, we have determined the difference between the time of detection in LVD and that of the kicker extraction signal at CERN, after accounting for the CERN-LVD baseline, i.e., $\delta T_{batch} = (t_{\nu} - t_{p}) - tof_{c}$. 
$\delta T_{batch}$ does not include yet the identification of the neutrino bunch originating the event. 
To check the detector operation, we have looked for coincidences in the whole LVD in a rather wide time window, namely $|\delta T_{batch}| < 100~ \mu s$. 
We have found 190 events in total, consistent with the $1.89\,\cdot\,10^{17}$ protons on target (p.o.t.) delivered during the bunched-beam run \cite {LVDmonitor}. \\
For the measurement of the neutrino velocity, we have used only events involving at least one of the Super-Set counters. 
We have found 79 of them out of 190, i.e., 40\% as expected (see previous section). 
Four of them are actually not usable for an accurate time-of-flight measurement: one has no complete information in the CERN database, while for three of them the ADC was malfunctioning. 
To limit the sources of systematic uncertainties, we have applied quality cuts to the remaining 75 events. First, because the number of photoelectrons for energy releases E$<$10 MeV is too low to guarantee that the counter is triggered by direct (i.e., fast) light, we have selected only events where E$>$10 MeV (ten excluded events). Second, we have required the non-saturation of ADC of the triggering counter, as the saturation would not allow a precise measurement of the $\delta_{LVD}$ term in (2) (seven excluded events). Finally on the remaining 58 events, we have tested different energy cuts, between 10 and 100 MeV. 
While the mean value of the time-of-flight distribution remains constant for any cut, the r.m.s. decreases as the threshold increases. 
It becomes stable for E$>$50 MeV, this value being our final choice for the energy cut. 
The described cuts reduce the sample to 48 events, that will be used for the final analysis. 
We note that among all the excluded events, there are five whose 
time-of-flight significantly deviates from the mean of the distribution of the original 79 events.
Those are five of the events for which E$<$10 MeV. Their number is consistent with that expected from background. Indeed, the background frequency of the Super-Set, mainly due to the rock radioactivity, is $f_{bk}\,=\,1.1\,$ s$^{-1}$. 
With $n\approx25000$ extractions, the expected number of background events is $N\,=\,n\,f_{bk}\,2 \cdot 10^{-4}\,\approx\,5.5$.\\
The 48 selected events  have been individually inspected: almost all of them show the presence of a muon track, as expected from a $\nu_{\mu}$ charged current interaction.
We show the distribution of the measured $\delta T_{batch}$ in figure \ref{fig:wfdistr} (black lines) where the sum of the associated digitized waveforms of the beam intensity monitor are also shown (grey lines).
The beam structure is clearly identifiable, and every LVD event can be associated to the closest beam-waveform peak.
For each event we then calculate the time difference, $\delta t$, with respect to the maximum intensity of the peak.
The distribution of $\delta t$ for the 48 events is shown in figure \ref{fig:finaldistr} (black histogram) compared with the superposition of all the peaks of the waveforms correlated to detected events (grey curve). 
The positive tail in the grey histogram is an artifact of the transfer function of the BCT system\cite{Pablo}. This effect does not influence our measurement since time calibrations are performed with respect to the position of the maximum.
The mean value of the measured distribution is:
\begin {equation} 
\delta t={\it tof}_\nu-{\it tof}_c =\,0.9\,\pm 0.6_{stat}~ {\rm ns}
%,\pm 4.1 (sys) \,ns.
\end{equation}
\begin{figure}[htb]
%\centering
\vspace{-.0cm}
\includegraphics[width=90mm] {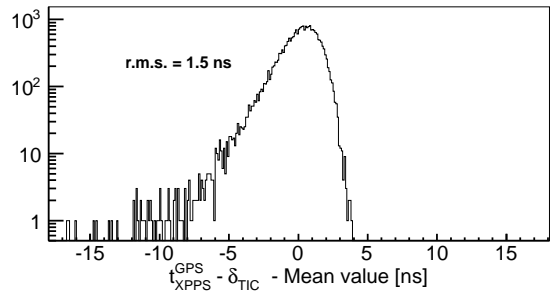}
\vspace{-.0cm}
\caption{System stability during the measurement: every 10 minutes the transit time of one Super-Set counter was measured.}
\label{fig:monitor}
\end{figure}
\begin{table}
\begin{ruledtabular}
%\begin{tabular}{|l|c|c|}
%Statistical uncertainties&ns&Error dist.\\
%\hline
%\hline
%PMT transit time correction&1.5&Gaussian\\
%GPS measurement&1.0& Gaussian\\
%Detector geometry &0.8& Gaussian \\
%Bunch width & 3.0& Gaussian \\
%\hline
%Total Statistical & 3.6& \\
%\hline
%\vspace{+0.5cm}\\
%\hline
%\end{tabular}
%\vspace{+0.5cm}\
%   \caption{Statistical  uncertainties on $tof_{\nu}$ measurement}
%    \label{table:tab_stat}
%\end{table}
%\end{ruledtabular}
%\end{table}
%\begin{table}
%\begin{ruledtabular}
\begin{tabular}{|l|c|c|}
Systematic uncertainties&ns&Error dist.\\
\hline
\hline
Baseline ($<$ 10 cm)& $<$~0.3~~~ & Gauss.\\
Const. corrections at CERN & 2.0 & Gauss.\\
BCT calibration & 1.0 & Gauss. \\
Time link calibration &1.1& Gauss.\\
GPS syncronization accuracy & 1.0 & Gauss.\\
PMT transit time correction&0.4&Gauss.\\
Absolute transit time calib. &1.5& Flat\\
Optical fiber length & 0.5 & Gauss.\\
Fluorescence lifetime & 0.6 & Exp.\\
%Light propagation &0.5& Flat \\
\hline
\hline
Total systematic & $\pm$ 3.2~~~ & \\
%\hline
%\hline
\end{tabular}
%\vspace{-0.2cm}
   \caption{Sources of systematic uncertainty in the measurement.}
    \label{table:tab_syst}
%\end{table}
\end{ruledtabular}
\end{table}
The long-term stability of the experimental setup has been monitored during the measurement by triggering  the LED in one of the counters of the Super-Set every 10 minutes and measuring the delay of the entire chain (see figure \ref{fig:monitor}).
The system, during the entire period, fluctuated with a r.m.s.=1.5 ns.
The asymmetry of the distribution is due to PMT pre-pulses \cite{prepulse}.\\
The systematic uncertainties associated to the measurement of $\delta t$ are summarized in Table \ref{table:tab_syst}. 
The baseline between LVD and CERN is known with an uncertainty lower than 0.3 ns. 
In fact, neutrino-induced muons responsible for events in the Super-Set can be generated as far as several hundreds of meters from LVD. 
This makes shorter the actual neutrinos baseline, while part of the distance is travelled by muons of different energy. 
This effect has been investigated through Monte Carlo and has a negligible impact on the measurement.
Constant delays at CERN give an additional uncertainty of 2.0 ns.
The uncertainty due to the system time inter-calibration \cite{Alessandro} has been measured and it is 1.1 ns. 
Also, the delay between the proton extraction time and the recording of the BCT waveforms by a digitizer has been measured, with an uncertainty of 1.0 ns. 
The variable correction $\delta_{LVD}$ (energy and counter dependent) introduces a further systematic uncertainty, 0.4 ns, evaluated through the propagation of the one associated to the energy measurement. 
Finally, the last three terms in Table \ref{table:tab_syst} are related to the constant corrections in the absolute time calibration of the Super-Set counters. 
They account for the unknown time delay between trigger and LED light generation, 2.6$\pm$1.5 ns; 
the uncertainty associated to the optical fiber length, 0.5 ns,
and the difference between the detector response to LED light and to particle ionization, $1.1 \pm 0.6$ ns (the scintillator decay time is $\tau$=3.32 ns \cite{Gd}).
By quadratically summing up all these contributions, we obtain a total systematic uncertainty of 3.2 ns.\\
{\it{Conclusions.-}} We have presented the measurement of the  neutrino velocity with the LVD experiment, through the detection of  muon-neutrinos from the CNGS beam. 
During 10 days of bunched-beam, LVD has detected 190 events in coincidence with the beam neutrinos. 
Among them, 79 have involved at least one counter in the Super-Set, an ensemble of 58 counters upgraded for precise timing measurements. 
To limit the sources of systematic uncertainties, we have applied quality cuts to these events. 
The resulting 48 have been used to determine the time-of-flight of $\nu_{\mu}$ with $<E>\,=\,$17 GeV on the CNGS baseline.
The deviation from the time expected from propagation at the speed of light has been found to be:
\begin {equation} 
\delta t=\,0.9\,\pm 0.6_{stat}\,\pm 3.2_{sys}~~ {\rm ns}
\end{equation}
The corresponding 99\% confidence limit on the speed of neutrino is: 
\begin{equation}
-3.8 \times 10^{-6} < (v_{\nu} - c)/c < 3.1 \times 10^{-6}~
%~ (99\% ~C.L.)
\end{equation}
These values are an order of magnitude lower than previous direct measurements.\\
By using the average neutrino beam energy, $<E_{\nu}>=17$ GeV, a limit is found on the relativistic mass of neutrino:
\begin{equation}
m_{\nu_{\mu}} < 47\,\, MeV/c^{2} \,\, (99\% ~C.L.) 
\end{equation}
{\it{Acknowledgements.-}} This  measurement was made  possible by the effort of G.~Di~Carlo, S.~Parlati and P.~Spinnato from the LNGS Computer Center;
G. Korga, from the Borexino Collaboration;
P.~Alvarez and J.~Serrano from the CERN staff;
 G.~Cerretto, V.~Pettiti and C.~Plantard from INRIM Torino;
Hector Esteban from the Real Instituto y Observatorio de la Armada (ROA), San Fernando, Spain and the INFN Torino Electronics Division.
We thank F. Vissani for discussions of great value to our work.\\
{\it{Note added in proof. -}} 
Recently we received a draft concerning a new measurement of the $\Delta_{const}^{CERN}$, i.e., the sum of constant delays at CERN \cite{Pablonote}.
The new values is: $\Delta_{const}^{CERN}\,=\,9522.4\pm 2.0$ ns.
By using it our result becomes:
\begin {equation*} 
\delta t=\,-0.3\,\pm 0.6_{stat}\,\pm 3.2_{sys}~~ {\rm ns}
\end{equation*}
The difference with respect to the previous result is well within our given uncertainties. 
The corresponding limit on the speed of neutrino, at 99\% C.L., becomes more stringent:
\begin{equation*}
-3.3 \times 10^{-6} < (v_{\nu} - c)/c < 3.5 \times 10^{-6}~
%~ (99\% ~C.L.)
\end{equation*}
and for the neutrino mass:
\begin{equation*}
m_{\nu_{\mu}} < 44\,\, MeV/c^{2} \,\, (99\% ~C.L.) 
\end{equation*}

\
\end{document}